\documentstyle[11pt,IAU207_pasp,twoside]{article}
\def\edcomment#1{\iffalse\marginpar{\raggedright\sl#1\/}\else\relax\fi}
\reversemarginpar

\begin{document}
\title{Clusters in extragalactic HII regions and their modelling}
 \author{Rosa Mar\'\i a Gonz\'alez Delgado}
\affil{Instituto de Astrof\'\i sica de Andaluc\'\i a (CSIC). Apdo. 3004, 
18080 Granada, Spain}
%\author{Ima Co-Author}
%\affil{The Name of My Institution, The Full Address of My Institution}

\begin{abstract}

This review deals with properties of very young ($\leq$ 10-20 Myr) stellar clusters 
which are in the nebular phase and are embedded 
in photoionized regions (classical extragalactic HII regions (RHIIs) or starburst galaxies).
Based on the analysis of the integrated light of these clusters at 
the UV and optical wavelengths, different techniques are discussed that allow to estimate in 
consistent the stellar content and the evolutionary state of the ionizing clusters.
Figures illustrating this contribution can be found in: http://www.iaa.csic.es/$\sim$rosa/

\end{abstract}

\section{Introduction}

RHIIs are amongst the brightest objects in galaxies. They have sizes up to a 
few 100 pc and their H$\alpha$ luminosity is larger than 10$^{39}$ erg s$^{-1}$
(Kennicutt 1984). The most luminous and brightest RHII in the Local Group is
30 Doradus in the LMC. It has a nebular size of 400 pc and an H$\alpha$ luminosity
10$^{40.2}$ erg s$^{-1}$,  produced by more than one thousand ionizing 
stars. However, most of the radiative and mechanical energy of the region is
provided by the central compact cluster R136. It contains more than one 
hundred stars within a few parcsecs, having a stellar density of $\sim$ 2 stars/pc$^2$
(Campbell et al 1992). This cluster is very young, 
$\sim$ 3 Myr (Vacca et al 1995). Its compactness contrasts with that of 
the second brightest RHII in the Local group, NGC 604 in M 33. As 30 Dor, 
NGC 604 is photoionized by a very young cluster (3 Myr old, Gonz\'alez Delgado \& P\'erez 2000) 
that contains a large population of O and B stars. However, it extends 
in a larger area with an effective radius of 60 pc (Hunter et al 1996; Drissen et al 1993).
Several subclumps are detected, but they have still a stellar density is which 
a factor 10 lower than R136. The stellar clusters in these two luminous and brightest 
RHIIs could represent two different modes of star formation: compact stellar clusters (called 
super star clusters SSCs) and a young extended population that may be responsible of
the diffuse UV light detected in more distant star forming regions. 

Starbursts show properties (nebular size, luminosity, ...)
that could be considered as scaled up versions of RHIIs where  
these two modes of star formation also co-exist (see Ma\'\i z-Apell\'aniz in this volume). 
Meurer et al (1995) have found that only 20$\%$ of the UV light of a sample of optically 
or UV selected starburst galaxies is provided by SSCs, and the remaining 80$\%$ could be
due to an extended unresolved population similar to the population in NGC 604.
In fact, Tremonti et al (2001) have shown that this diffuse UV light detected in 
the starburst galaxy NGC 5253 is not scattered light from the SSCs but it originates 
from unresolved, more extended stellar clusters. 

Most of the RHIIs and starbursts are far-away enough that the main characteristics
of the stellar clusters can be estimated only through the analysis
of the integrated light. UV light is potentially very useful to estimate the properties of these stellar
clusters because it represents direct emission from hot massive stars. 
Therefore, UV light provides direct evidence of the location 
and properties of the most recent stellar clusters in a star forming galaxy. On the 
other hand, these massive stars are the main source of the ionization, mechanical
heating and chemical enrichment of the interstellar gas. The spectrum of the 
ionized gas is dominated by nebular emission. Here, I will show how to unveil
the properties of young stellar clusters embedded in photoionized regions through the 
modelling of their UV and optical light. 

\section{UV-light}

The UV spectra of RHIIs and starbursts are mainly dominated by absorption lines (Rosa et al 1984;
Kinney et al 1993). They form in the wind and photosphere of massive
stars. However, these resonance lines can also form in the interstellar medium
of the star forming regions. Instead of strong absorption lines, some RHIIs 
show a few or weak nebular lines (see for example NGC 2363A, Drissen et al 2000). 
An inventory of the most important absorption lines are:  

\begin{itemize}

\item{Wind lines: The strongest one are OVI $\lambda$1036, NV $\lambda$ 1240, SiIV $\lambda$1400, 
CIV $\lambda$1550, HeII $\lambda$ 1640, NIV $\lambda$1720. They show a P-Cygni profile and/or 
are shifted by $\sim$ 1000-3000 km s$^{-1}$. These lines depend
on stellar luminosity, mass loss rate, terminal velocity and age of the stars 
(Leitherer \& Lamers 1991). Therefore, the shape of the line profiles 
in the UV integrated light of a stellar population is related to its content in massive stars.
Thus, these features can be used to constrain the properties (such as age, and the slope and upper 
mass limit of the IMF) of the stellar clusters}

\item{Photospheric lines: they are much weaker than the wind lines. SV $\lambda$1502,
 FeV $\lambda$1430, SiIII $\lambda$1417, 
FeV $\lambda$1360-1380, OV $\lambda$1371 are produced by O and early B stars; and 
SiII $\lambda$1265, 1485, SiIII triplet ($\lambda$1295-1300), CII $\lambda$1335 and 
CII $\lambda$1247, 1427 also form in late type B stars. At longer wavelengths, 
FeII $\lambda$2570-2615 and MgII $\lambda$2780-2825 features also form. These lines are
also useful to constrain the age of the stellar cluster, but also the metallicity. 
In particular blends of these lines in 1360-1380 and 1415-1435 show a strong dependence with Z
(Leitherer et al 2001).  }

\item{The most important interstellar lines in the $\sim$1000-1800 \AA\ spectral range are:   
SiII $\lambda$1260, OI+SiII $\lambda$1302, CII $\lambda$1335, SiIV $\lambda$1400,
SiII $\lambda$1526, CIV $\lambda$1550, FeII $\lambda$1608, AlII $\lambda$1670. Low ionization
lines are very useful to study the kinematics of the ionized gas because the 
interstellar component usually dominates the stellar contribution (Gonz\'alez Delgado et al 1998a). 
They are also useful to derive the metallicity of the gas 
(Pettini et al 2000). The high ionization interstellar lines are blended with the wind lines, a careful 
separation between both components is necessary. However, when the stellar cluster is young (2-8 Myr),
 wind lines dominate over the interstellar ones. }

\end{itemize}

Evolutionary synthesis models that predict the wind and photospheric lines have been 
proposed by Mass-Hesse \& Kunth (1991), Robert et al (1993), Leitherer et al (1995a), 
Gonz\'alez Delgado et al (1997b), de Mello et al (2000). These models have been extensively 
applied to date stellar clusters in RHIIs and Starbursts observed by IUE and HST (Table 1). 
Most recent results are commented in section $2.2$. 

\subsection{Estimating the masses of stellar clusters}

The mass of the stellar cluster can be estimated comparing the UV luminosity provided
 by the cluster with the predictions by evolutionary synthesis models. However, 
the mass estimated represents a lower limit to the virial mass of the cluster
because the UV light does not account for low mass stars. On the other hand, these 
estimates depend on the extinction correction applied to the observed UV flux. 
For young clusters not heavily obscured that emit a significant fraction of their 
bolometric luminosity in the UV, the extinction can be estimated comparing the 
observed with the UV intrinsic spectral distribution predicted 
by the evolutionary synthesis models, because the UV spectral slope (F$_{UV}
\sim {\lambda}^{\beta}$) is independent of the IMF and the star formation history.
It is correlated with the color excess E(B-V) estimated from the Balmer decrement
and with L$_{FIR}$/L$_{UV}$ (Meurer et al 1997).
This indicates that some UV photons are absorbed by dust and re-emited at FIR wavelengths.
This is the prediction of a dust screen model. However, variations on this geometry can 
also explain this correlation (Charlot \& Fall 2000). In addition, the UV luminosity depends
on the extinction laws used to do the correction. Several attenuation laws have
been proposed (Rosa \& Benvenuti 1994; Mas-Hesse \& Kunth 1999; Calzetti 1997) 
to be applied to star forming regions, but all of them are coincident in showing a 
2175 \AA\ bump weaker than in the galatic extinction law.

Masses estimated for ionizing stellar clusters are given in Table 1. The masses of the 
clusters in RHIIs are 
$\sim$ 10$^4$ M$\odot$, much lower than the mass estimated for starbursts 
($\sim$ 10$^6$ M$\odot$). The latter  
probably correspond to several clusters that contribute to a larger aperture.

\begin{table}
\caption{Age and mass estimated for the clusters responsible of the photoionization 
in RHIIs and Starbursts}
\begin{tabular}{lccccl}
            &      &                  &          &      & \\
Name        & Age  & Mass             & Aperture &      & Authors\\
            & Myr  & 10$^5$ M$\odot$  & pc       &      &         \\
            &      &                  &          &      & \\
30Dor       & 3    & 0.2              & 14       & IUE  & Vacca et al 1995 \\
NGC 2363B   & 2.5  & 0.1              & 13       & STIS & Drissen et al 2000 \\
NGC 604     & 3    & 0.5              & 60       & IUE  & Gonz\'alez Delgado \& P\'erez 2000 \\
RHIIs M101  & 3    &                  &          & FOS  & Rosa \& Benvenuti 1994 \\
NGC 5253    & 1-8  & 0.01-0.4         & $\leq$10 & STIS & Tremonti et al 2001 \\
NGC 4214    & 4-5  & 1                & 20       & FOS  & Leitherer et al 1996 \\
Antenae     & 2-8  & 1                &          & GHRS & Whittmore et al 1999 \\
He2-10      & 4    & 20               & 75       & GHRS & Johnson et al 1999 \\
NGC 3049    & 4    & 10               & 60       & STIS & Gonz\'alez Delgado \& Leitherer 2001 \\
NGC 1741    & 4    & 30               & 400      & GHRS & Conti et al 1997 \\
NGC 7714    & 4.5  & 50               & 350      & GHRS & Gonz\'alez Delgado et al 1999a \\
Seyfert 2   & 3-6  & 30-500           & 300      & GHRS & Gonz\'alez Delgado et al 1998b \\
Darwf Gal.  & 2-13 & 0.1-1000         &$\geq$1000& IUE  & Mass-Hesse \& Kunth 1999 \\
Nuclear SB  & 3-6  & 100-400          &$\geq$1000& HUT  & Gonz\'alez Delgado et al 1998a \\

\end{tabular}
\end{table}

\subsection{General Results}

The main results of these studies can be summarized as follows: UV integrated light 
is characterized by an instantaneous burst a few Myr old, populated
by a Salpeter IMF with stars up to M$_{up}$ $\geq$ 50 M$\odot$. 
When the integrated light is emitted by extended areas ($\sim$ 100 pc), the UV spectra
are equally well fitted by continuous star formation lasting for a few Myr. These results
indicate that clusters form with a very small age spread. In fact, 
this is the case for the starburst He2-10 (Johnson et al 2000) 
in which the clusters are chained along $\sim$ 90 pc with a mean separation $\leq$ 10 pc,
and they are all 4-5 Myr old. 
However, a good exception is the case of the low metallicity RHII NGC 2363.
It contains two clusters (called A and B) within a distance of $\sim$ 70 pc. The UV 
spectra of the two clusters have a very different morphology. Cluster A shows only
a few weak interstellar lines plus CIV $\lambda$1550 in emission, but the spectrum of
cluster B shows strong wind lines (Drissen et al 2000). These differences are due to an age 
spread in the formation of the two clusters. Cluster A is still embedded in dust (age 
$\leq$ 1 Myr), however, cluster B is more evolved as indicated by the strong wind CIV 
line detected, suggesting an age of 2.5 Myr. These differences in the age could be due
to several bursts of star formation in a time scale of 10 Myr over a spatial
scale of 400 pc, probably triggered by the pass of a small satellite galaxy.

There are indications that the IMF and the global star formation processes are the 
same in metal rich clusters as they are in metal poor ones. 
A good example is the metal-rich, barred starburst
NGC 3049. HST observations done with STIS/MAMA (FUV) indicate that most of the UV light 
is emitted by the central 2 arcsec. The wind lines detected in the spectrum indicate that 
the cluster(s) in the inner 50 pc form in an instantaneous burst 3-4 Myr ago. Even though 
the metallicity of the stars is oversolar, stars more massive than 50 M$\odot$ form
in the cluster(s). However, field extended stellar population could form with a different
IMF than the clusters. This is the case of the low metallicity starburst NGC 5253. Tremonti 
et al (2001) have obtained STIS/MAMA (FUV) narrow slit spectra of 8 stellar clusters plus 
several inter-cluster regions of diffuse light. They find that the UV light of all clusters is
well fitted by instantaneous bursts with ages between 1 and 8 Myr that follow a Salpeter IMF
extending up to 100 M$\odot$. However, the field spectrum is better fitted by continuous star
formation models with either M$_{up}$=30 M$\odot$ or an IMF slope steeper than Salpeter. However,
other more sophisticated explanations are possible involving age and disruption effects in time
scale of $\sim$ 10 Myr.

\subsection{Uncertainties in the models}

The basic ingredients in the evolutionary synthesis models are: evolutionary stellar tracks, 
stellar atmospheres and stellar libraries used to predict the stellar parameters. Thus, the results 
obtained depend on these inputs to the evolutionary code. Here, I comment the limitations and 
improvements of the models concerning to the UV stellar libraries, evolutionary stellar tracks
and the parametrization of the IMF.  

\subsubsection{UV stellar libraries}

 Most of these evolutionary synthesis models use as input a stellar
library built with the UV spectra of stars in the solar neighborhood. Then, it is very
suitable to describe clusters with metallicity close to solar. Because, wind lines 
depend on mass loss rate (in consequence on metallicity), the predictions of these
models could be no adequate for low metallicty stellar clusters. Recently, 
Leitherer et al (2001) have built a UV stellar library with HST observations of 
O3 to B0 stars from the LMC and SMC. They provide new evolutionary synthesis models at 
1/4 Z$\odot$. These models predict photospheric lines much weaker, as expected
from lower elemental abundances. However, the behaviour of the wind stellar lines
is more complex; while, NV  $\lambda$1240 and SiIV $\lambda$1400 do not scale 
monotonically with metallicity, CIV $\lambda$1550 is significantly affected, showing a weaker 
P Cygni profile. Thus, while the wind NV and SiIV may be equally well predicted using the 
solar stellar library, CIV and the photospheric lines are overpredicted in low metallicity 
clusters, inducing a wrong estimation of the age and of the IMF parameters. 

\subsubsection{Evolutionary stellar tracks}

Massive stars are fast rotators. Rotation induces instabilities that produce transport
of angular momentum and chemical elements to the outer radiative envelope. When the effect
of the rotation is taken into account, the evolution of massive stars is significantly 
modified. New stellar tracks with rotation from the Geneva group (Maeder \& Meynet 2000)
predict with respect to the non-rotating models: a) Massive stars in the Main Sequence 
are bluer. b) The Wolf-Rayet phase is longer and starts earlier in the evolution. 
c) The ratio of the number of Wolf-Rayet to O type stars is larger, and Wolf-Rayet C to
Wolf-Rayet N is lower. 
None of the evolutionary codes developed until now include these new tracks. 

\subsubsection{Stochastic effects on the IMF}

Evolutionary synthesis models use continuous functions (usually power laws) to distribute 
the number of stars formed in a cluster, and thus, they do not reproduce the discontinuous 
nature of star formation. This difference between the nature and the models has an important
effect, in particular, in stellar clusters with small number of stars. Cervi\~no et al (2000)
have performed Montecarlo realizations for clusters of several masses (10$^3$ M$\odot$,
10$^4$ M$\odot$ and 10$^5$ M$\odot$) to simulate the stochastic nature of the IMF. They 
find that the properties of the stellar cluster are not affected if the mass of the cluster
is $\geq$ 10$^5$ M$\odot$. But, for lower masses, the widths of the parameter distributions
compared with analytical values are proportional to the mass transformed into stars, 
assumptions on IMF and age. Because the mass estimated for RHIIs is lower than 10$^5$ M$\odot$,
stochastic effects on the IMF is an important uncertainty in the predictions of the properties
of these stellar clusters.

\section{The optical-IR light}

In contrast to the UV light, the optical to infrared spectrum of RHIIs is dominated by the 
nebular emission lines. The interstellar gas is photoionized by Lyman continuum photons 
provided by the most massive and hot stars in the clusters. Gas cools down via recombination and 
collisionally excited nebular lines. However, when the stellar cluster 
is $\sim$ 3-5 Myr old, broad emission lines (HeII $\lambda$4686, CIV $\lambda$5800 bumps) 
from Wolf-Rayets are detected in the optical spectra. Evolutionary synthesis models that make
predictions of the equivalent width and luminosity of these features are in Schaerer \& Vacca (1998),
Starburst99 (Leitherer et al 1999), Cervi\~no et al (2001). 

Other optical stellar features are
the high order Balmer series and some HeI lines in absorption. They form in the photosphere 
of O, B and A stars. These stellar features are normally overwhelmed by the nebular contribution 
if the cluster is in the nebular phase. However, if the stars and gas do not have the same spatial 
distribution, then,  
H$\delta$, H$\epsilon$, H8, H9 ,... and HeI $\lambda$4471, HeI $\lambda$4026, HeI $\lambda$3819,
can be detected in absorption (Gonz\'alez Delgado \& P\'erez 2000). The strength and profile
of these lines show a strong dependency with the age of the stellar cluster, and they are 
very useful diagnostics if the cluster is older than $\sim$ 5 Myr old (Gonz\'alez Delgado et al 1999b). 
In addition to the Balmer and HeI absorption lines, the spectra of young clusters in the post-nebular 
phase (10-20 Myr old) show also the CaII triplet at 8550 \AA\ and CO $\lambda$2.2 $\micron$m
band. Beautiful examples of clusters in this phase are most nearby SSCs (called A and B) in 
the irregular galaxy NGC 1569 (Gonz\'alez Delgado et al 1997a; see Gilbert et al in this volume and 
the most recent works based on HST by Origlia et al 2001 and Maoz et al 2001). Evolutionary
synthesis models treating these lines include Mayya (1997), Garc\'\i a-Vargas et al (1998) and
Origlia et al (1999). 

In the following sections, I discuss only the results obtained from the modelling of the nebular 
lines.    

\subsection{Modelling of the nebular emission lines}

The modelling of the nebular 
lines is a degenerated problem because of the dependence on radiation field 
from the ionizing stellar cluster, the chemical abundance of the gas, and the density 
structure and geometry of the region. Even so, it is possible to constrain the stellar content,
age and IMF of the cluster by coupling an evolutionary synthesis and a photoionization code.
Observable quantities to constrain the models are:

\begin{itemize}
\item{Collisional excited lines: Forbidden to Balmer line ratios, 
as [OIII] $\lambda$5007/H$\beta$; $\eta$ parameter, defined as 
([OIII] $\lambda$5007/[OII] $\lambda$3727)/([SIII] $\lambda$9069+9531/[SII] $\lambda$6717+6732)  
sensible to the effective temperature of the cluster (V\'\i lchez \& Pagel); low 
ionization lines like [OI] $\lambda$6300/H$\beta$; and  infrared fine structure lines such as 
[NeIII]/[NeII],[ArIII]/[ArII] or [SIV]/[SIII]}

\item{Recombination lines: The ratio of ionized He to ionized H, like HeI $\lambda$5876/H$\beta$, 
depends on the ratio of He to H ionizing photons, Q(He)/Q(H); thus, it is sensitive to 
the effective temperature of the stars if T$_{eff}\leq$ 40000 K (Stasi\'nska 1996)}.

\item{Hydrogen recombination lines equivalent width, e.g. Ew(H$\beta$), depend on Q(H) 
and also on the optical continuum luminosity emitted by the cluster. It is a good indicator of 
the age of the stellar cluster.}
\end{itemize}

Unfortunately these observational constrains also depend on other quantities like chemical
abundances, differential extinction, and the contribution of other more evolved stellar populations.
The uncertainties associated to these constrains and with the models are commented in section 3.2.
A detailed modelling can be done follow the illustrative flow chart proposed by Garc\'\i a-Vargas 
et al (1997). A photoionization code such as CLOUDY requires in addition to the spectral energy
distribution (SED) from the evolutionary synthesis code the following inputs: a) Chemical abundance 
of the gas estimated from the emission line ratios. b) Gas electron density estimated from the 
[SII] $\lambda$6717/[SII] $\lambda$6732 ratio. c) Assumption about the geometry of the gas. If
a sphere is assumed, the inner and outer radius has to be specified, that can be estimated from
high spatial resolution H$\alpha$ images. d) Total ionizing photon luminosity, that it can be estimated
from total H$\alpha$ flux or from UV continuum luminosity provided by the stellar clusters.

The general results from the most recent grids of models (e.g. Stasi\'nska et al 2001; 
Kewley et al 2001; Moy et al 2001; Dopita et al
2000; Bresolin et al 1999), 
tailored models at optical wavelengths 
(e.g. Luridiana \& Peimbert 2001 for NGC 5461;
Gonz\'alez Delgado \& P\'erez 2000 for NGC 604; Stasi\'nska \& Schaerer 1999 for IZw18; 
Gonz\'alez Delgado et al 1999a for the nuclear starburst of NGC 7714; Luridiana et al 1999 for NGC 2363; 
Garc\'\i a-Vargas et al 1997 for the RHIIs of NGC 7714) 
and at infrared wavelengths 
(e.g. Colbert et al 1999 for Arp 299; Forster-Schreiber et al 1999 for
M82; Schaerer \& Stasi\'nska 1999 for NGC 5253 and IIZw 40)
can be summarized as: a) Stellar clusters are very young and well described
by an instantaneous burst. b) Clusters form following a Salpeter IMF, but in high metallicity RHIIs 
M$_{up}$ is $\leq$ 40-30 M$\odot$. It seems to contradict the results obtained at the UV light and 
at the optical based on the detection of Wolf-Rayet features in high metallicity starbursts (Schaerer et al 2000).

\subsection{Critical points and uncertainties in the modelling}

Most of these models fail to reproduce the Ew(H$\beta$) and 
some lines ratios like [OIII] $\lambda$4363/H$\beta$ and [OI] $\lambda$6300/H$\beta$ (see e.g. Stasi\'nska
2000). However, the input to the photoionization models 
(e.g. the hypothesis that the nebula is radiation bounded; constant electron density 
along the RHII) and the ingredients used in the evolutionary synthesis models 
(in addition to those mentioned in section 2.3, the stellar atmospheres assumed to built the SED) 
have an important impact on the results of the photoionization models. Changes in these 
ingredients help to explain the divergences between the observed quantities and the predictions.

\subsubsection{Stellar atmospheres}

It is not clear how realistic is the hardness of the SED predicted by evolutionary synthesis models, 
in particular in the Wolf-Rayet phase. Most of the evolutionary codes use the Kurucz (1993) stellar atmospheres.
However, Stasi\'nska et al (2001) have produced a grid of photoionization models using the COSTAR 
stellar atmospheres (Schaerer \& de Koter 1997) and they find an important impact on Q(He)/Q(H), 
produced by the effect of the wind of O main sequence stars in the models.
Starburst99 (Leitherer et al 1999) which is
optimized for predictions for young clusters, uses the non-blanketed Schmutz et al (1992) model 
atmospheres for stars with strong winds. These models predict a large increase in the 
nebular excitation at the onset of the Wolf-Rayet phase due to a large energy above the HeII ionization
limit. In consequence, high-metallicity models predict much higher excitation than observed in 
metal-rich RHIIs (Bresolin et al 1999; Dopita et al 2000; Kewley et al 2001) due to a large number of 
Wolf-Rayet stars. It is expected that line blanketed extended atmosphere models should produced a
softer spectrum above 4 Ryd but it is not clear how the SED could be from 1 to 4 Ryd (Schaerer 2001). These line blanketed
stellar atmospheres should predict more similar emission line ratios to those observed in 
high metallicity RHIIs. Note that the present models predict older ages and a lower number of massive stars,
thus, a steeper IMF and a lower M$_{up}$, than the real values in metal-rich stellar clusters. This may be the
reason of the discrepancy between the results obtained analyzing the nebular lines and the 
the modelling of UV wind lines in metal rich clusters.

\subsubsection{Density structure}

Most models assume that the electron density is constant. However, there are evidences that 
RHIIs can have a density structure (P\'erez et al 2001; Casta\~neda et al 1992). Luridiana \&
Peimbert (2001) have modelled the RHII NGC 5461 in M101. They find that an asymmetric nebula with a
gaussian electron density distribution powered by a young cluster $\sim$3 Myr old reproduces most
of the constraints. They find that the results strongly depend on the assumed density law. If it is
constant, then the models overestimate the hardness of the ionizing field, and in consequence, it 
gives an erroneous determination of the age and IMF of the stellar cluster.

\subsubsection{Radiation bounded hypothesis}

Most of the photoionization models assume that the nebula is radiation bounded. Consistent models 
using the UV light and nebular emission lines (e.g. Gonz\'alez Delgado et al 1999a) find that 
the mass of the cluster estimated from the UV continuum is in agreement with the value estimated 
with the total Balmer recombination lines (e.g. H$\beta$). This result indicates that the leakage
of ionizing photons is very small and, in general, that the radiation bounded hypothesis is correct. However,
a small leakage in some direction is possible (Leitherer et al 1995b; Heckman et al 2001). Photon
leakage affects preferentially the highest frequencies. These photons reach further into the
gas, and filaments or diffuse gas can absorb them. Thus, larger values of 
[OI] $\lambda$6300/H$\beta$ can be predicted by photoionization models if there is some leakage 
of ionizing photons that are absorbed by filaments (e.g. in IZw18 Stasi\'nska \& Schaerer 1999). 
This result rules out shocks as
additional heating source that is able to explain larger values of [OI] $\lambda$6300/H$\beta$.

Photon leakage has also been proposed to explain the low values of Ew(H$\beta$) compared with 
the prediction of the models. Moy et al (2001) compute photoionization models for several covering
factors (cf). However, to explain the Ew(H$\beta$) observed in HII galaxies a cf=0.1 is required.
This means that 90$\%$ of the ionizing photons have to escape from the nebula. Thus, leakage of
ionizing photons is not a good explanation to the low values of Ew(H$\beta$) observed in
HII and starburst galaxies. Other alternatives, such as differential extinction between gas and stars
or the contribution to the continuum of an underlying old stellar population have been suggested
(Raiman et al 2000). 

\section{Conclusions}

Very general conclusions from the studies discussed here are:

\begin{itemize}

\item{The modelling of the wind and photospheric UV lines and the nebular Optical-IR lines
are powerful tools to estimate the stellar content, IMF and age of embedded stellar clusters 
in RHIIs. However, we have to be aware that the results depend on the ingredients of the 
stellar evolutionary tracks and stellar atmosphere models, in particular on Wolf-Rayet stars.}

\item{Stellar clusters are well described by an instantaneous burst of a few Myr old. 
The stellar masses distributed with a Salpeter IMF with M$_{up}$ $\geq$ 50 M$\odot$. Lower M$_{up}$ estimated in high 
metallicity RHIIs could be related to the use of inadequate stellar atmosphere models to compute 
the SED.}

\end{itemize} 

\acknowledgements I thank to the organizers, in particular to Eva Grebel and Dough
Geisler for inviting me to give this talk and for their financial support. I am
also very grateful to Christy Tremonti for sending me a draft of her paper on NGC 5253
in advance of publication, and to Grazyna Stasi\'nska, Daniel Schaerer and Enrique P\'erez for their comments
from a thorough reading of the paper.

\section*{Discussion}

{\it Sakhibov:\, }
Did you consider a case of continuous star formation? How do you account
for the fraction of the Lyman continuum photons which do not contribute 
to the ionization process, missed leakage?
\\ 
\noindent
{\it  Gonz\'alez Delgado:\, }
UV light may be equally well fitted by instantaneous
or continuous star formation lasting for only 3-5 Myr. However, the nebular emission lines 
of RHIIs are better modelled by instantaneous bursts. Continuous star formation models predict
larger excitation and Ew(H$\beta$) than the values observed. Leakage of ionizing photons 
can be accounted computing photoionization models with covering
factor (cf) less than 1. Moderate changes (cf lower but close to 1) do not alter significantly 
the strongest emission line ratios. Lower values of cf are not 
realistic because the mass of the clusters estimated from the nebular recombination lines is in 
good agreement with estimations done using the UV continuum luminosity.  
\\ 

{\it D'Antona:\, }
Do the upper mass limits you gave for a few systems include the new rotating 
stellar models of the Geneva group?
\\ 
\noindent
{\it  Gonz\'alez Delgado:\, }
Many of the results presented here were analyzed with Starburst99. This code 
has implemented the new stellar tracks of the Geneva group {\em without} rotation (Meynet et al 1994;
Schaerer et al 1993 and Charbonnel et al 1993; Schaller et al 1992). To my knowledge
no evolutionary synthesis code has yet implemented the new stellar tracks with
rotation. 
\\ 

{\it Elmegreen:\, }
Your suggestion of two modes of star formation is consistent with optical 
observations, but in the infrared, most embedded stars in the solar neighborhood
are forming in clusters, and in this sense, there is only one mode: it is all
clustered. So what you see optically, is perhaps more a difference in survival
and dispersal of clusters than in modes of formation of stars.
\\ 
\noindent
{\it  Gonz\'alez Delgado:\, }
It could be true that diffuse UV emission detected in starbursts could be produced by 
stellar clusters that have been disrupted in a time scale of 10-20 Myr. Tremonti et al
(2001) have proposed this interpretation for the starburst galaxy NGC 5253.
\\

\end{document}